\documentclass[12pt,letterpaper]{amsart}
\usepackage{amsmath}
\usepackage{setspace}
\doublespace
\usepackage[ansinew]{inputenc}

\usepackage{fixltx2e}
\usepackage{textcomp}
\usepackage{fullpage}
\usepackage{amsfonts}
\usepackage{verbatim}
\usepackage[english]{babel}
\usepackage{pifont}
\usepackage{color}
\usepackage{lscape}
\usepackage{indentfirst}
\usepackage[normalem]{ulem}
\usepackage{booktabs}
\usepackage{natbib}
\usepackage{float}
\usepackage{latexsym}
\usepackage{url}
\usepackage{hyperref}
\usepackage{epsfig}
\usepackage{graphicx}
\usepackage{amssymb}
\usepackage{bm}
\usepackage{array}
\usepackage{mhchem}
\usepackage{ifthen}
\usepackage{caption}
\usepackage{hyperref}
\usepackage{amsthm}
\usepackage{amstext}
\usepackage{subfig}
\usepackage{yfonts}
\usepackage{epic}
\usepackage{latexsym}
\usepackage{epsfig}
\usepackage[mathscr]{euscript}
\usepackage[all]{xy}

\raggedright
\renewcommand{\section}[1]{%
\bigskip
\begin{center}
\begin{Large}
\normalfont\scshape #1
\medskip
\end{Large}
\end{center}}

\renewcommand{\subsection}[1]{%
\bigskip
\begin{center}
\begin{large}
\normalfont\itshape #1
\end{large}
\end{center}}

\renewcommand{\subsubsection}[1]{%
\vspace{2ex}
\noindent
\textit{#1.}---}

\renewcommand{\tableofcontents}{}

\bibpunct{(}{)}{;}{a}{}{,}  

\newtheorem{thm}{Theorem}

\DeclareMathAlphabet{\mathpzc}{OT1}{pzc}{m}{it}

\newtheorem{lemma}{Lemma}
\newtheorem{theorem}[lemma]{Theorem}

\newtheorem{proposition}[thm]{Proposition}  

\newcommand{\EE}{{\mathbb E}}
\newcommand{\PP}{{\mathbb P}}

\vspace{1.5in}

\begin{document}
\begin{flushright}
Version dated: \today
\end{flushright}
\bigskip
\noindent Compatibility and saturated data

\bigskip
\medskip
\begin{center}

\noindent{\Large \bf Circumstances in which parsimony but not compatibility will be provably misleading}
\bigskip



\noindent {\normalsize \sc Robert W. Scotland$^1$, and Mike Steel$^{2}$}\\
\noindent {\small \it 
$^1$ Department of Plant Sciences, Oxford University, Oxford OX1 3RB, UK\\
$^2$ Biomathematics Research Centre, University of Canterbury, Christchurch, New Zealand} \\
\end{center}
\medskip
\noindent{\bf Corresponding author:} Mike Steel, Mathematics and Statistics, University of Canterbury, Christchurch, New Zealand E-mail: mike.steel@canterbury.ac.nz\\
Fax: +64-3-3642587; Phone: +64-21329705


\vspace{1in}

\newpage

\subsubsection{Abstract}   \\
Phylogenetic methods typically rely on an appropriate model of how data evolved in order to infer an accurate phylogenetic tree. 
For molecular data, standard statistical methods have provided an effective strategy for extracting phylogenetic information from aligned sequence data when each site (character) is subject to a common process.  However, for other types of data (e.g. morphological data),  characters can be too ambiguous, homoplastic  or saturated to develop models that are effective at capturing the underlying process of change. To address this, we examine the properties of a classic but neglected method for inferring splits in an underlying tree, namely, maximum compatibility. By adopting a simple and extreme model in which each character either fits perfectly on some tree, or is entirely random (but it is not known which class any character belongs to) we are able to derive exact and explicit formulae regarding the performance of  maximum compatibility. We show that this method  is able to identify a set of non-trivial homoplasy-free characters, when the number $n$ of taxa is large, even when the number of random characters is large.  By contrast, we show that a method that makes more uniform use of all the data --- maximum parsimony ---  can provably estimate trees in which {\em none} of the original  homoplasy-free characters support splits.  \newpage

\section{Introduction}

Inferring phylogeny is a central goal for systematics because nested sets of monophyletic taxa provide a pivotal anchor point for the construction of classifications \citep{apg98} as well as for understanding evolutionary history \citep{fel04}.  During the last 20 years, monophyletic taxa have been predominantly estimated using model-based inference methods and molecular sequence data. The continuing role of morphological data to estimate monophyletic taxa has not been without discussion and controversy  (e.g. \cite{hill87, sco03, jen04, wie04}) 
but the number of morphological analyses in comparison to analyses using DNA sequence data continues to decline as judged by the number of morphological matrices deposited in TreeBASE  \citep{pie10}.

The exact role of morphological data for phylogenetic inference, and in particular morphological data from fossil taxa, has been a particular source of debate \citep{pat81, gau88, sco03, gra04, wie04, spr08, wie10}.   These views occupy the complete spectrum from the opinion that fossils are best interpreted in the light of monophyletic taxa based on extant organisms (Patterson 1981) to the view that a combined total evidence approach  utilizing all data is to be preferred for inferring phylogeny \citep{hue96, nix96, wie10}. Other solutions for integrating morphological data in phylogenetic inference  involve the analysis of both molecular and morphological data separately to seek congruence among data sets on the basis that this provides the strongest evidence that phylogenetic reconstruction is accurate \citep{pen86, sal13, swo91}.   Still other approaches recommend using molecular scaffolds in which trees derived from molecular sequences are used to constrain the analysis of the morphological data on the basis that morphological characters may contain too much homoplasy or saturated, non-independent signal \citep{spr08, dav14}. 
Despite these differences of opinion surrounding morphological data and phylogeny reconstruction,  most authors agree on the importance of morphological data from both extant and fossil taxa to provide a full and comprehensive understanding of evolutionary history.

Despite misgivings and legitimate concerns about morphological data possibly being saturated, too homoplastic and/or non-independent 
\citep{wag00, spr07, spr08, dav14},  these same authors remain committed to finding solutions to include at least some morphological data in estimates of phylogeny. This is relevant because of the important role of taxon sampling in phylogeny reconstruction \citep{hill96, hill98} combined with the fact that most taxa that ever lived are now extinct and therefore only exist as fossils \citep{hill87}.  Furthermore, how to best integrate and assign fossil taxa for dating nodes of phylogenetic trees is also a topic of some interest \citep{ron12}. 
 In this context, we revisit compatibility as a method for estimating monophyletic taxa in the context of morphological data. Our motivation stems from the claim that morphological datasets often contain ambiguous, saturated phylogenetic signal that can approximate to random data for extant \citep{kel14}  as well as fossil taxa  \citep{wag00,  spr07, spr08, dav14}.
 We focus on compatibility, a method that seeks to discover unique compatible characters describing splits in the underlying phylogenetic tree, rather than attempting to explain or model all congruent and incongruent characters \citep{far83, lew01}.

Almost fifty years ago,  \citet{wil65} and \citet{cam65} (and later \citet{leq69, est72, leq72, leq75, est77, far77, fel78, est79, mea81, fel82, mea85}) explored compatibility -- termed `character consistency' by \citet{wil65} and `character congruence' by \citet{pat82} -- as a method for analyzing morphological data to infer phylogeny.  Despite some more recent discussion (e.g. \cite{dep91, wil94, dre97, fel04, gup07}) compatibility has remained on the periphery of methods for inferring phylogeny, as it has largely been set aside, initially in favour of maximum parsimony, and, more recently, by model-based methods for inferring  phylogeny from DNA sequence data.  For an overview of 
compatibility methods in phylogeny reconstruction see \cite{mea85} or \cite{fel04}.

Compatibility was described by \citet{wil65} as a method of character `weighting' based on the phylogenetic significance of the character. \citet{wil65} aimed to capture taxonomic procedure explicitly in a new, more rigorous way by using a method that weighted unique unreversed character states that were consistent between each other and a hierarchical hypothesis, to the exclusion of more `fickle'  character states. We here interpret `fickle characters' {\em sensu} \cite{wil65} as referring to homoplasy but also other factors (including analogy, inaccurate character concepts, inaccurate coding) that can lead to effectively random patterns of character states shared between taxa.

Compatibility was therefore a form of character weighting, seeking to give maximum weight to characters that evolve once and display no homoplasy. \cite{leq69} stated that a character is compatible with a tree if it can evolve on that tree without homoplasy. He stated that a character with $N$ states that requires $N-1$ changes on a tree, is compatible with that tree. He further reasoned that the best tree was the tree that maximized the total number of compatible characters \citep{leq69}. In a sense, the method of character compatibility formalized a phylogenetic method that captured the intuitive taxonomic practice of recognizing taxa based on conserved non-homoplastic characters \citep{wil65}.  The justification provided in the compatibility literature for attempting  to identify and utilize only characters that evolve once in estimates of phylogeny were three-fold.  First, a set of characters that exhibit a higher level of compatibility than would be expected by  chance alone may reflect a common  process, namely descent with modification \citep{wil65}.  Second, this statistical property (compatibility) does not apply to more noisy or homoplastic characters and therefore these should be excluded from further consideration \citep{leq69, leq72, leq75, mea81}.  Third, that the history of taxonomy and the recognition of natural groups  has hitherto utilized non-homoplastic characters and that inferences about convergent and homoplastic characters were after the fact interpretations from classifications based on compatible characters \citep{wil65, pat82}.

Compatibility methods have been most often compared and  contrasted with cladistic parsimony methods that were developed and refined during a similar time period \citep{hen66, far77, far79, far83}. 
Compatibility and parsimony methods were shown to share certain characteristics including statistical inconsistency \citep{fel78}.  In contrast, the two methods -- parsimony and compatibility -- differed most fundamentally in their treatment of homoplasy and character conflict \citep{wil65, far79, mea85, dep91}. 

 Compatibility methods seek to infer phylogeny from uniquely derived non-homoplastic  characters that are consistent and non-random in their distribution \citep{wil65} whereas parsimony methods seek to explain all characters by incorporating and minimizing {\em ad hoc} assumptions of homoplasy \citep{far83}.  To account for character conflict, parsimony analysis adjusts the level of universality of some  characters to fit the general most parsimonious tree of all characters, such that all characters are treated as informative, and the principle of parsimony determines the most economical (fewest number of changes) explanation of the data 
 \citep{far83, dep91}. Comparing parsimony and compatibility directly, \cite{dep91} concluded that compatibility ``has little power in determining hypotheses of character evolution in the presence of incongruence". 
The view of \cite{dep91}  is that parsimony analysis seeks to explain all data relative to a model of character evolution whereas, in contrast, compatibility
seeks to discover compatible characters only  \citep{wil65}.  

The role and utility of character weighting for inferring phylogeny was explicitly explored in the phylogenetics literature  during this period \citep{wil65, far69, nef86, car88, sha89, hil91, hil93}.   Adherents of parsimony also explored weighting schemes to give reduced  weight to characters based on empirically determined levels of homoplasy  \citep{far69, car88, gol}.    The relative importance and weight ascribed to various characters within a dataset remains an active topic of discussion in contemporary phylogenetics 
 \citep{edd04, fel04, lem09, cox14}, and several factors have been investigated, including:  strong signal to noise ratios \citep{sal13}, random data \citep{wen99}, saturation \citep{wag00, kel14}, 
codon bias \citep{cox14},  third position changes \citep{cox14}, reliability and homoplasy \citep{gol}, heterogeneity of substitution rates among different lineages \citep{ho09}, differences between DNA and protein data \citep{kum08}, the use of BLOSUM62 matrix for aligning proteins \citep{edd04}, and the weighting of transitions versus transversions \citep{pos01}.   We therefore consider a re-examination of compatibility methods as part of a much wider research agenda seeking to explore and model the relative strength of phylogenetic signal within datasets  \citep{gol, sal13, cox14}.

More specifically, our motivation to re-examine compatibility stems in part from  a recent meta-analysis of morphological datasets in TreeBASE  \citep{kel14} demonstrating  that many morphological datasets contain very little signal when compared with random data. Similar results were observed in paleontology for 48 out of 56 fossil datasets \cite{wag00} in a study that identified a deterioration of phylogenetic structure through time due to character state exhaustion (saturation) in many clades. These observations -- very few uniquely derived morphological characters (synapomorphy) and much morphological data
that is very problematic -- may explain a historical paradox in systematics, i.e. that morphological data has been extraordinarily successful at estimating monophyletic taxa by synapomorphy  albeit for a limited number of nodes but at the same time most morphological data are inherently problematic for inferring phylogeny \citep{wag00, sco03, gra04, spr07, spr08, wag12, bap13, bap14, dav14}.

Today, maximum likelihood and Bayesian techniques are the main tools for inferring phylogenetic trees from most (genetic) character data \citep{fel04}. Although a great variety of stochastic models have been developed and applied for aligned DNA sequence site data, there has been comparatively much less work on modelling the evolution of  discrete morphological characters. One exception is \cite{lew01}, who showed how a symmetric Markovian model with a finite number of states could be applied to morphological data in a maximum likelihood setting (see also  \cite{hue08} for a somewhat different Bayesian analysis).  An obstacle for many types of morphological (or fossil) data is that, in contrast to DNA site substitutions, there is unlikely to be a common mechanism across the characters (e.g. ratios of `branch lengths' within a tree may vary across characters) and the absolute rates of evolution may also vary in unknown ways
across the characters.   For example, some characters may be highly conserved, with just a single `innovation' occurring once in an evolutionary tree, while other characters may have flipped states many times, resulting in an essentially random pattern of states at the tips of the tree.

In this paper, we analyse what happens when the data arise in precisely such a manner:  some characters evolve without homoplasy, while others are essentially random, but we have no idea which class a given character belongs to. We adopt this  scenario because we can obtain exact results, and it shows what is possible in the extreme. Nevertheless  it also provides some guidance on what may be expected in less polarised settings.    
Our approach complements the study by \cite{sus05} who showed how biases affect standard tree reconstruction methods in the presence of varying
degrees of randomisation within sequences or sites.

One can model  such extreme data using standard and simple Markov models on an phylogenetic tree,  such as the Mk model of \citep{lew01}, as follows:  The characters that evolve without homoplasy simply correspond to characters that are evolving at a very slow rate (i.e. they are highly conserved)  while those that are essentially random are evolving at a very fast rate.   This  model can be viewed either as a two-fold mixture of the common mechanism model (i.e. the ratio of the branch lengths within a tree  is the same across all characters, with these branch lengths merely being scaled up or down according to whether the character is conserved or random)  or as a model in which there is no such constraint on the branch length ratios.   To keep matters simple, we consider binary characters (which corresponds to $k=2$ in the Mk model), but similar results could be developed more generally.

Although evolution at a low substitution rate will lead to (mostly) homoplasy-free characters, many of these characters will be unvaried (i.e.  all taxa would be in the same state).
However,  as noted by \cite{lew01}, such unvaried characters are generally not of interest in comparative morphological studies, and it is generally more relevant to consider data that excludes such
unvaried characters.  This `censoring' of the data (effectively by the investigator looking for characters that reveal differences between taxa) provides no problem for our analysis, and we will accommodate this additional  perspective explicitly in Theorem~\ref{mainthm}.

\section{Definitions:  Character compatibility and  random binary characters}

We begin with some definitions.    A {\em binary character} $\chi$ on a set $X$ of $n$ taxa is an assignment of a state (0 or 1) to each taxon $x$ in $X$.  We say that $\chi$ is {\em non-trivial}
if at least two taxa are in one state and at least two taxa are in a different state; otherwise, the character is {\em trivial}.  Thus, a trivial character is either unvaried (all taxa receive the same state) or it is an `autapomorphy' (one taxon receives one state, and  all remaining taxa the other state). Non-trivial characters are also sometimes referred to as `(parsimony) informative' characters.   The {\em partition} of a character $\chi$ 
refers to the partition of $X$ into (at most) two parts that $\chi$ induces (i.e. for each partition, there are two characters of that partition, obtained by interchanging the states 0 and 1). 

A phylogenetic $X$-tree {\em displays} a binary character if the character fits on the tree with one state change at most (i.e. no homoplasy).    For example, in Fig. 1, the character that assigns taxa $a$, $b$ and $e$ one state and the remaining taxa a different state is displayed by the tree on the right, but not by the tree on the left. Two characters are {\em compatible} if there is a phylogenetic $X$-tree that displays both characters.  This is equivalent to requiring that the two subsets of taxa (one for each character) that have a state that is different to the state of some arbitrary reference taxon $x_0 \in X$ comprise a pair of sets  that are either disjoint (i.e. have empty intersection)  or nested (i.e. one set contains the other).  A  character is compatible with all other possible characters if and only if that character is trivial.

More generally, a sequence of binary characters is compatible if there is a phylogenetic $X$-tree that displays them all. It is a classic result that a set of characters is compatible if and only if every pair is. Moreover, there is a unique minimally-resolved phylogenetic tree that displays these characters, where the non-trivial splits of the tree correspond to the bipartitions of $X$ induced by the non-trivial characters (see, e.g. \cite{sem03}).   

Given a character $\chi$ on $X$, and a phylogenetic $X$-tree, $\chi$ is {\em compatible with $T$}  precisely if $T$ or some resolution of $T$ displays $\chi$. An example is shown in Fig. 1.

 \begin{figure}[h]
\begin{center}
\resizebox{12cm}{!}{
\includegraphics{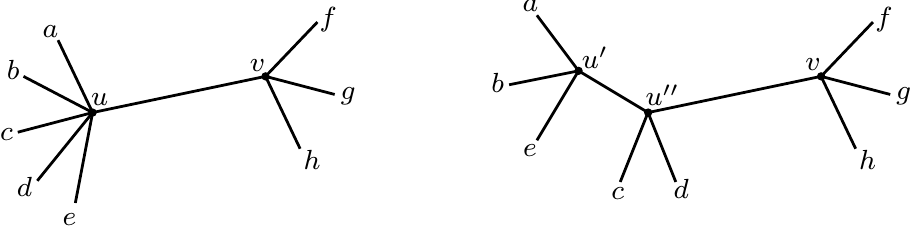}
}
\caption{The tree on the right resolves the tree on the left by the addition of an edge ($u',u''$) to resolve vertex $u$. The tree on the right displays any character $\chi$ that assigns
 taxa $a,b,e$ one state and $c,d,f,g,h$ a second state,  so $\chi$ is compatible with this tree as well as  the tree on the left (even though that tree does not display $\chi$).}
\label{fig1}
\end{center}
\end{figure}

A {\em random binary character} $f$ on a set $X$ of $n$ taxa, is an assignment of a state (0 or 1) to each taxon $x$ in $X$, performed independently across the taxa and
with an equal probability of 0 or 1 for each taxon. Thus a random binary characters simply picks one of the  possible $2^n$ binary characters on $X$ with equal probability.  

Note that a  two-state symmetric Markov process on a phylogenetic tree produces a random binary character in the limit as the rate of substitution across all (or sufficiently many) branches
grows. Such  processes were investigated in the context of morphological evolution in \cite{lew01}.   Here, we work with the limiting value of completely random data since it allows
exact calculations. However, our results  have a bearing when the data contains characters that are `near random' (i.e. near saturation), as we describe briefly in the concluding comments. 

We will henceforth regard phylogenetic trees as unrooted (but there is no real loss of generality with this assumption).

\section{Results}

The main result in this paper is the following. 

\begin{theorem}
\label{mainthm}

Suppose  a data-set $D$ consists of a sequence of binary characters (with or without any unvaried characters removed), that consists of a sequence $S_1$ of compatible  characters and a sequence $S_2$ of $M$ independent random binary characters, all on the same set $X$ of $n$ taxa (these sequences are intertwined and so we are not told which character belongs to which class).

\begin{itemize}
\item[(i)]
With a probability of at least $1-\epsilon$ where $\epsilon = 2M^2 \left(\frac{3}{4}\right)^n$, no two characters in $S_2$ are compatible. Consequently, with probability at least $1-\epsilon$ the following hold:
\begin{itemize}
\item[(1)]
there exists a maximal compatible subset $S$ of $D$ that contains all of $S_1$; 
\item[(2)] any such $S$ falls into one of the following two cases:
(a)  $S=S_1$; or (b) $S$ consists of all (or all but one) characters from $S_1$ and one character from $S_2$; 
\item[(3)]
a maximal compatible sequence $S$ of characters for $D$ can be identified by an efficient (polynomial time) algorithm from $D$ (without knowing in advance which characters are in $S_1$ and $S_2$).
\end{itemize}

\item[(ii)]
If, in addition,  $M \cdot\left(K \left(\frac{1}{2}\right)^{n-3} + \left(\frac{1}{2}\right)^{K}\right) \leq \delta$,
where $K$ is the number of distinct partitions produced by non-trivial characters in $S_1$, then with a probability of at least $1-\delta$, Case (b) in Part (2) of (i) will never arise.

\item[(iii)]
Let $m$ be the number of non-trivial characters in $S_1$ and let $L$ be the average value (over those $m$ characters) of the smaller number of taxa in the split determined by the character. 
Let $T_1$ be any binary tree that displays all of the compatible characters in $S_1$.  If  $M$ and $n$ are both large (e.g. $>30$) and  $27(L-1)^2m^2/2Mn< 1$, then the expected number of phylogenetic trees that are more parsimonious for $D$ than $T_1$ and yet display {\bf none} of the $m$ perfectly compatible non-trivial characters in $S_1$ grows exponentially with 
$n$ (it is at least $10^n$ for $n\geq 30$).\end{itemize}
\end{theorem}

The proof of this theorem is provided in the Appendix, with one small exception: here we outline the simple algorithm referred to in the third claim of part (i).  Such an algorithm is relevant, because, although software is available to search for maximum compatible subsets of characters \citep{fel93}, in general it is an intractable  (NP-hard) problem to find a largest subsequence of compatible characters in an arbitrary collection of binary characters.  This was established by \citep{day} who reduced the maximum compatibility problem for binary characters to the well-known problem of finding a maximum clique (a set of vertices all connected to each other) in an arbitrary graph.   This connection is the same as we use here -- given a sequence $D$ of binary characters construct a graph that has a vertex for each character in $D$,  with an edge between two vertices if the corresponding characters are compatible.  Now, for $D$ partitioned (as here) into two disjoint sets -- $S_1$ and  $S_2$ -- if no two of the (random) binary characters in $S_2$ are compatible, and each pair of binary characters in $S_1$ is compatible, the resulting graph is easily seen to be a `chordal' graph (i.e. every cycle of length four or more has a chord).   For such graphs, there exist fast (i.e. polynomial-time) algorithms for finding maximum cliques, based on constructing a `perfect elimination ordering' for the graph (for details, see \cite{gav}). Moreover, this also suggests a  useful diagnostic for testing whether the model described here is appropriate: simply check whether the associated character compatibility graph is chordal (a process that can also be carried out quickly \citep{ros76}).  

The reader should be clear that Theorem 1 describes a prediction of a model, and for real data there will generally be greater ambiguity as to the identity of the set of homoplasy-free characters than the sharp results described in parts (i) and (ii) of that theorem provide.
We now turn to some examples and graphs  to illustrate the content of Theorem~\ref{mainthm}.

\subsection{Examples}

\begin{itemize}
\item[(1)]
As a simple application, suppose we have 30 taxa, and 20 binary characters, of which 8 are distinct, non-trivial and perfectly compatible, while  the remaining 12 are random.
Then applying Parts (i) and (ii) with $n=30, K=8, M=12$ (noting that  $2(12)^2(3/4)^{30} = .051$, and $12 \cdot\left(8 \left(\frac{1}{2}\right)^{27} + \left(\frac{1}{2}\right)^{8}\right) = .047$) the probability the eight perfectly compatible characters comprise the unique maximum compatible subset of characters for this dataset is at least 90\%. In this example, $M$ is not large enough to  apply part (iii) usefully.

\item[(2)]  Suppose that we have $n=100$ taxa, and suppose our data $D$ is made up of a sequence $S_1$ of perfectly compatible non-trivial binary characters interspersed (in some unknown way) with a sequence $S_2$ of $M$ random binary characters.   Provided that $M$ is no more than (say) 10,000, then the conditions for Theorem~\ref{mainthm}(i) applies with $\epsilon =0.0001$, since
$2M^2 \left(\frac{3}{4}\right)^n  \leq 0.000064$. Therefore  we can easily distinguish/find the perfectly compatible characters from $S_1$ in $D$, and any maximum compatible tree for $D$ will display all or all but one of these characters and, at most, one other (random) character. Also,  at least one maximum compatible tree for $D$ contains every character in $S_1$.

If, in addition, at least 20 of the 97 possible non-trivial character partitions are present in $S_1$ ( i.e. $K \geq 20$), then the condition for Theorem~\ref{mainthm}(2) applies with
$\delta = 0.01$. If we combine this with the previous result, it is $\sim$99\% probable that a maximal compatible subsequence of characters from $D$ will consist of exactly just the characters in $S_1$.

Finally,  to illustrate part (iii) of Theorem~\ref{mainthm}, suppose that $S_1$ contains at most 30 non-trivial characters in total, counting repetitions (i.e. $m\leq 30$), with $L=7$, and that  $M=5000$, say.  
Then, $n$ and $M$ are sufficiently large, and $\frac{27(L-1)^2m^2}{2Mn} \leq 0.88$, which is small enough to  apply part (iii).
Thus,  the expected number of binary trees  that are simultaneously (a) more parsimonious (on the entire data set) than $T_1$ (any given binary tree that 
displays all of the perfectly-compatible characters in $S_1$), and (b) that fail to display any of the non-trivial characters in $S_1$, is at least $10^{100}$. 
This suggests we should have no confidence that a maximum parsimony tree for the entire data set  would have any success in displaying any (let alone several) of the perfectly compatible informative characters in $S_1$.  Moreover, provided that $K\geq 20$, we saw in the previous paragraph that the maximal compatible sequence of characters from $S$ is almost certain to consist of just the characters in $S_1$.

\end{itemize}

\bigskip

\subsection{Graphs}

The above examples provide a `snapshot' of how the results of Theorem~\ref{mainthm} apply. However, it is also helpful to visualise the interplay of the
various parameters ($n, K, M, m$) that result from the inequalities in Parts (i)--(iii).  In Fig. 2, the graphs (A), (B) and (C) illustrate aspects of  Parts (i), (ii) and (iii) of Theorem~\ref{mainthm},  respectively.

 \begin{figure}[h]
\begin{center}
\resizebox{16cm}{!}{
\includegraphics{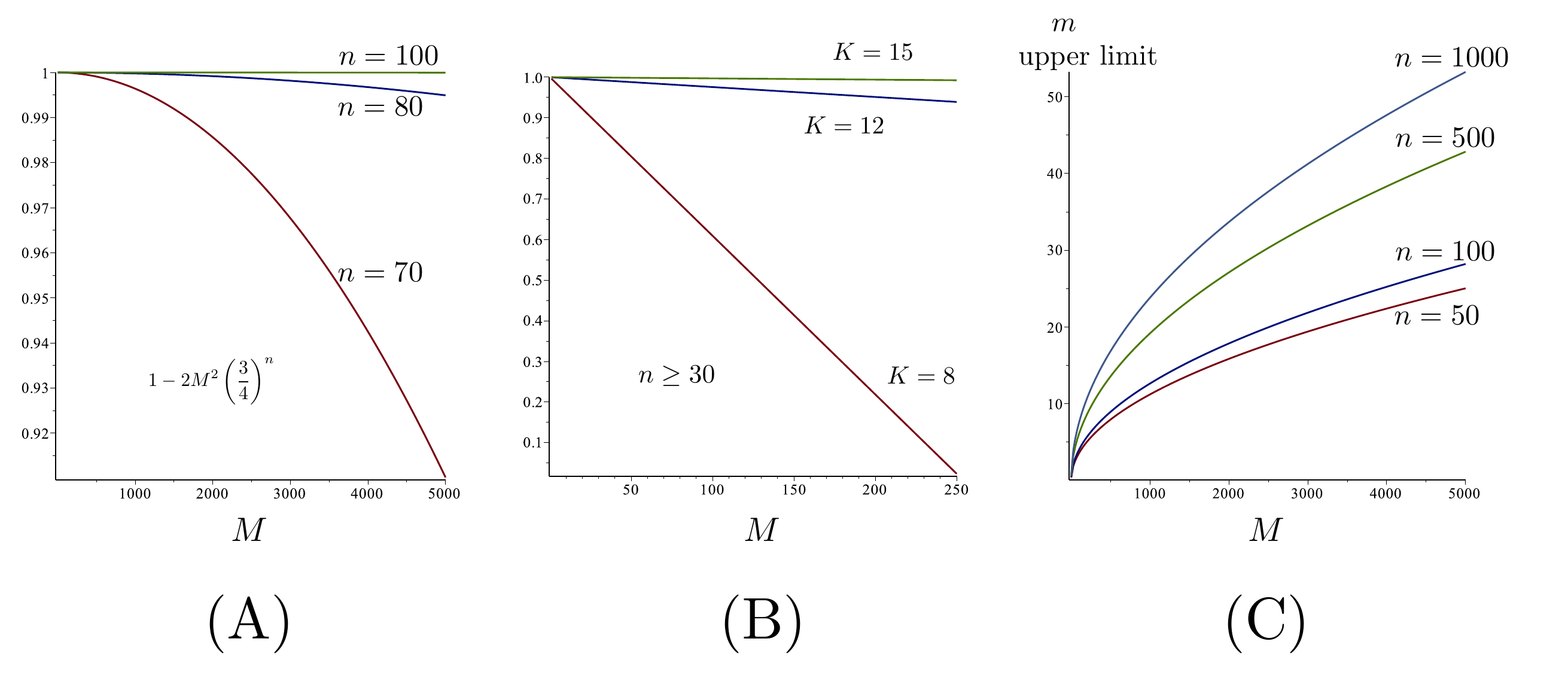}
}
\caption{(A) The probability that no two characters from among $M$ random binary characters are compatible lies above the curves shown.
(B) The probability that none of $M$ random binary characters on 30 or more taxa are compatible with a sequence of non-trivial binary characters that induce $K$ distinct splits lies above the curve shown (for $K=8, 12, 15$). 
(C) The maximal value of $m$ allowed in Theorem~\ref{mainthm} so that  when $m$ nontrivial binary characters are sampled at random from a Yule pure-birth tree $T$, a large $(>10^{n}$) number of trees are more parsimonious than $T$ and yet display none of the $m$ characters. }
\label{fig2}
\end{center}
\end{figure}

Graph (A) shows the graph of $1-2M^2 \left(\frac{3}{4}\right)^n$ verses $M$, which is a lower bound on the probability that amongst $M$ random binary characters, no pair of characters are compatible.  While this bound decays quadratically with $M$, the exponential dependence on $n$ ensures that when the number of taxa is large, this probability will be close to 1 even for relatively large values of $M$.

Graph (B) shows a lower bound on another probability -- namely that none of $M$ random binary characters are compatible with 
a sequence of binary characters that induce $K$ non-trivial splits. In this case the value of $n$ plays a vanishing role once it becomes more than about  15 or so;  we have drawn the graph for $n=30$ but it would look identical for any larger value of $n$.

Graph (C) shows the upper bound on the number $m$ of compatible nontrivial binary characters as a function of $M$ for the number of nontrivial binary characters for which maximum parsimony performs as described in Theorem~\ref{mainthm}(iii). In this graph we have estimated $L$ by considering the expected size of clades in Yule trees up to size $n/2$ (using the (asymptotic) estimate $2\ln(n/2)$ which follows from  Theorem 4.4 of \cite{ros06}).  Notice that the upper bound on $m$ is quite modest as a function $M$ (the number of random binary characters), and grows relatively slowly with $n$.  In other words, to achieve this extreme performance of maximum parsimony, the characters in $S_1$ will form a tree that is only partially resolved (this may not be necessary, but the current proof of part (iii) allows only sub-linear growth of $m$ with $n$).

\newpage

\section{Discussion}

Methods for inferring phylogenetic trees from character data  are most often an exercise in modelling all data. In these situations, all data are viewed as providing evidence of phylogeny and the task is to infer character-state changes at the correct node(s) on a tree. In these circumstances some character-state changes are unique whilst others contain homoplasy.   A confounding factor for phylogenetic inference are high rates of homoplasy that approximate to random data, such that phylogenetic signal is masked. In addition, mistakes in DNA sequence alignment, ambiguous coding regimes and disagreements about the correct conceptualization of morphological structures can add a level of non-evolutionary noise to a data matrix.  In this sense, data matrices contain unique signal (characters that contain no homoplasy), homoplastic signal (characters that change more than once but provide evidence for phylogeny at various nodes on a tree) and noise (characters that contain saturated levels of homoplasy with no phylogenetic signal and also characters that do not accurately reflect or capture evolutionary history e.g. mistakes in coding and conceptualization of morphological data).  Therefore, here,  `noise' refers to 
saturated levels of homoplasy and other factors (analogy, inaccurate character concepts, inaccurate coding, character dependency, continuous rather than discrete data, alignment error etc) that lead to essentially  random patterns of character states shared between taxa \citep{wen99}. When the data exhibit little homoplasy and noise, then evolutionary trees can be reliably reconstructed from simple methods, such as maximum parsimony and neighbour joining with dense taxon sampling   \citep{gra98, hill96}.  For data where the characters exhibit moderate but widely varying degrees of homoplasy (across the tree), according to some common process across the characters, then simple methods like maximum parsimony can be misleading, depending on the branch lengths of the true tree \citep{fel78}, while maximum likelihood and Bayesian approaches have more sound robust properties (e.g. consistency) for recovering the correct tree from sufficient data.  In both these cases, maximum compatibility tends to perform poorly alongside the standard methods that  use all of the available data more even-handedly. 

However, when the data consist of a combination of some characters that exhibit very little noise alongside other  characters that are very noisy (and we do not know which class a given character belongs to),  a standard method that regards all characters as having the same signal-to-ratio strength  is problematic. In this case, our results suggest that a more accurate strategy is to find a maximum compatible subset of characters, as this will contain (nearly all) the characters that exhibit very little noise and, provided the number of taxa is large,  very few additional characters from the highly noisy class.  Essentially, the separation of `signal' and `noise' is achieved by the characters themselves;  maximum compatibility merely identifies which class each character belongs to. The accuracy of this approach relies on a simple mathematical fact:  the probability that two random binary characters are compatible with each other converges to zero exponentially quickly as the number of taxa grows.  Consequently, in this setting, the mutually compatible characters are likely to directly reflect and capture the common underlying process of descent with modification, or, in some circumstances, concerted convergent evolution \citep{hol10}. 
A further advantage of compatibility is that it is not as sensitive to reduced taxon sampling as other methods that seek to make more uniform use of data. In contrast, for methods that seek to explain high rates of homoplasy at the correct nodes on a tree, dense  taxon sampling is crucial for accurate reconstruction \citep{gra98}.

We should stress that we are not advocating a wholesale return to maximum compatibility; as mentioned, there are settings where it is clearly inferior to parsimony and/or likelihood and Bayesian approaches.  Instead we argue that maximum compatibility may have some advantages  when data are such that each character either has very low or very high associated noise. Note that maximum compatibility may lead to an unresolved tree with polytomies (unlike most other tree reconstruction methods which tend to return a fully-resolved tree). 
Moreover, just because a clique of compatible characters is significantly larger than expected for random characters, these compatible characters may fit a tree that is different from the species phylogeny if those characters have been subject to convergent evolution \citep{hol10}.

Note that a highly noisy character (under our symmetric two-state model) will tend to have about half the taxa in one state and half in the other.  However, the observation of such  near-equality in the character state counts does not, in itself, mean that the character is noisy;  after all, completely homoplasy-free evolution can also produce such a character when a change of state occurs near some central edge of a tree that has around half the taxa on one side, and half on the other.

Our results show that a large number of  `noisy' characters will overwhelm the signal for parsimony present in the `noise-free' characters if they are all treated on a level footing as samples from an identical process. By  contrast,  our results show that maximum compatibility is accurate even when the number of random characters $M$ grows exponentially with the number $n$ of taxa (provided $M$ grows no faster  than $(4/3)^{n/2}$). Also, although our analysis deals with an extreme model (characters with zero or infinite noise), standard continuity arguments imply  that our results will provide some indication of how maximum compatibility will behave in less extreme settings.

Our analysis of maximum parsimony was chosen for simplicity, but it has implications for other phylogenetic tree reconstruction methods such as maximum likelihood (ML). To see this, suppose the data consists not just of compatible and random binary characters, but also many unvaried characters. Data of this type would be generated by a 2-rate model - very slow rates (leading to mostly unvaried but also some  homoplasy-free characters), and very fast rates (leading to random characters).  Theorem 7 of \cite{tuf97} implies that the maximum likelihood tree for this data analysed under a model that assumes a common mechanism and a constant rate across characters is always a maximum parsimony tree for this data, when the proportion of unvaried characters is sufficiently large.  This means that Theorem~\ref{mainthm} will apply in this setting (with maximum parsimony replaced by ML) since the unvaried characters play no role in either parsimony or compatibility analyses.   Our use of maximum parsimony was also confined to the simplest form of this method in which all character state changes are weighted equally, and all characters are also given equal weight; relaxing the second of these constraints has been explored by various authors, including for morphological data \citep{gol}.

We  restricted our analysis to binary characters, as these enjoy the property that a collection of them is compatible if and only if every pair are. The results could, in principle, be extended to multi-state characters; however, the mathematical analysis would be more complex, since pairwise compatibility is only a necessary (but not sufficient) condition for the compatibility of a collection of such characters. Within the two-state model one can also consider moving from a symmetric to a non symmetric model of substitution. In this case the probability that two random characters
are compatible increases, though it still decays exponentially with $n$ (for any fixed probability that a given taxon is in state 1).  A further minor extension of our results is that the requirement that the random binary characters be independent in Parts (i) and (ii) of Theorem~\ref{mainthm} can be weakened to requiring just pairwise independence,
since the proof of both parts relies only on Boole's inequality.

We mention a further caveat; Theorem~\ref{mainthm}(iii) refers to the expected number of trees that are more parsimonious that $T_1$ and yet fail to display any non-trivial character in $S_1$.
 The proof (Appendix) relies on showing that a randomly selected binary tree has a positive probability of being more parsimonious than $T_1$ (and also a  high chance of failing to display any non-trivial characters in $S_1$).  This suggests that the chance of $T_1$ being a (globally) most-parsimonious tree for the data should be infinitesimal (given that there are $>10^{180}$ trees on 100 taxa, and each has a reasonable chance of `beating' $T_1$).  However, some care is needed here: if we consider the event that $T'$  is more parsimonious  for the data than $T_1$, then these events (one for each $T'$)  are not independent.  This lack of independence does not  cause any problem in the statement of Theorem~\ref{mainthm}(iii), as it  refers to expectation, and the expectation of a sum of random variables is the sum of the expectations, regardless of whether they are dependent or not.  While it seems reasonable to expect that $T_1$ would be very unlikely to be a maximum parsimony tree for the data, further analysis would be needed to formally prove this.

The motivation for exploring a method that seeks to identify monophyletic taxa from a small subset of discrete character data will be obvious to most systematists who have studied patterns of variation across a clade of any size. Put simply, clear-cut discrete characters are few and far between. It is uncontroversial to state that morphological data and evolutionary novel character concepts (e.g., carpels, nucleic acid, seeds, legumes, vertebrae, amnion etc.) have been developed and refined hand in-hand, alongside the context of monophyly and classification (e.g., carpels of angiosperms, nucleic acids of life, seeds of spermatophytes, legumes of leguminosae, vertebrae of vertebrates, amnion of amniotes). These character concepts were discovered by a combination of good observational skills coupled with a subsequent hypothesis of monophyly that could be examined in the context of other hypotheses of monophyly by a process of reciprocal illumination throughout the history of systematics. In agreement with \cite{wil65}, we suggest that compatibility is a method that captures these aspects of systematic practice and is therefore worthy of consideration for inferring monophyly from morphological data.

Compatibility offers a further way to analyse morphological data independent of molecular data. For those who wish to analyse morphological data in combination with molecular data then compatibility can be implemented as a distinct data partition within a total evidence context. Compatibility could be used to screen out random signal and identify compatible characters for subsequent analyses in combination with molecular data.  In a sense, compatibility is less ambitious and perhaps less attractive than methods that seek to model all available data but at the same time it is, for some classes of data, more realistic in accepting a lack of resolution \citep{bap13} and a limited number of `good' morphological characters.  It is also possible that compatibility offers a new perspective for the study of morphological character evolution that attempts to incorporate two basic empirical findings of systematics. One,  that there are very few unreversed conserved characters and, two, that the majority of characters are problematic to model as discrete data as their phylogenetic distribution approaches saturation.

Here we have shown that for a model in which each character either fits perfectly on some tree, or is entirely random (but it is not known which class any character belongs to) we are able to derive exact and explicit formulae regarding the performance of maximum compatibility. The is significant because it is  perhaps the first time that any tree reconstruction method (on any number of taxa) can be analysed so exactly under a model that involves randomness in the data.  Furthermore, we show that compatibility is able to identify a set of non-trivial
homoplasy-free characters, when the number $n$ of taxa is large, even when the number of random characters is large. This is significant because one might have expected that, by chance alone, compatibilities within a few of the random characters would result in a number of incorrect splits being estimated by compatibility, and we provide precise conditions under which this will not occur. By contrast, we show that a method that makes more uniform
use of all the data -- maximum parsimony -- can provably estimate trees in which none of the original homoplasy-free characters appear as splits. This is significant because  `by chance alone' the random data can overwhelm the phylogenetic  signal in the homoplasy-free characters through the eyes of some methods (eg. parsimony) but not others (e.g. compatibility). 
While compatibility excludes much of the data, and so may result in unresolved trees, this conservative feature of the method has an advantage in the extreme model we study of being relatively immune to the influence of the random characters when the number of taxa is large.  On the other hand, maximum parsimony -- or indeed maximum likelihood when many additional constant characters are present -- is influenced 
by all  of the characters, so a large number of random ones will tend to lead to trees that display none of the homoplasy-free characters as splits. 

Taken together our results are significant for contemporary systematics because, although they deal with an extreme model, the  mathematical results provide a caution that in this setting `more can be less' -- methods that attempt to score a tree using all the characters may  miss most (or indeed all) of the unique  non-trivial characters that have evolved without homoplasy. Yet there exist other methods (such as compatibility) that can be immune to this, at the price of being more conservative.  In future work it would be useful to explore the extent to which this holds under less extreme models, where any mathematical analysis would be much less straightforward.

\section{Funding}
MS thanks the NZ Marsden Fund and the Allan Wilson Centre for helping fund this research.  RWS thanks the Leverhulme trust for  funding this research.

\section{Acknowledgments}
We  thank the reviewers, editor and associate editor for several helpful suggestions.

\bibliographystyle{plainnat}
\bibliography{scotland_steel}

\newpage






\section{Appendix: Proof of main result}

\subsection{Preliminaries}

Suppose we have an unrooted phylogenetic $X$-tree $T$, which may or may not have polytomies.  Our first result provides a bound on the probability that a random binary character
is compatible with $T$, and an exact expression for the probability that two random binary characters are compatible with each other.  
Part (b) of the following proposition differs from an earlier result by \citet{mea81} (based on the even earlier work of \citet{wil65})  who considered the probability that a pair of random binary characters are compatible, conditional on the number of taxa in each state for the two characters.

\begin{proposition}
\label{helps}
\mbox{ } 
\begin{itemize}
\item[(a)]  The probability that a random binary character is compatible with a given binary character that divides $X$ into blocks of size $r$, and $n-r$ is:
$$\left(\frac{1}{2}\right)^{r-1} + \left(\frac{1}{2}\right)^{n-r-1} - \left(\frac{1}{2}\right)^{n-2},$$
whenever  $1 \leq r \leq n-1$ (the probability equals $1$ if $r \in \{0,1,n-1, n\}$). 
\item[(b)]
The probability $p_n$  that two random binary characters on a set of size $n$ are compatible is given by:

$$p_n = 4 \left(\frac{3}{4}\right)^n - \left(\frac{1}{2}\right)^{n-1}\left(3-\left(\frac{1}{2}\right)^{n-1}\right).$$
Thus, $p_n$ is bounded above by  $4 \left(\frac{3}{4}\right)^n$ (and is asymptotically equivalent to it as $n$ grows).

\item[(c)]
For an unrooted phylogenetic tree on a leaf set $X$ of size $n$, the probability $p_T$ that a random character on $X$ is compatible with $T$ satisfies: 
$$ p_T \leq \sum_{ v\in I(T)} \left(\frac{1}{2}\right)^{n-{\rm deg}(v)},$$
where $I(T)$ is the set of interior (internal) vertices of $T$ and ${\rm deg}(v)$ is the degree of vertex $v$ (i.e. the number of edges incident with $v$).

\item[(d)] Moreover, for any unrooted phylogenetic tree $T$ that has at least $k$ interior edges, and $n$ leaves, we have the upper bound:  $$p_T \leq k \left(\frac{1}{2}\right)^{n-3} + \left(\frac{1}{2}\right)^{k}. $$

\end{itemize}
\end{proposition}

Before turning to the proof of this result (which is central to the proof of Theorem~\ref{mainthm}) we illustrate its application.

{\bf Example:}  To illustrate Part (b) of Proposition~\ref{helps}, consider $n=3$ and $n=4$ (for which $p_3=1$ and $p_4 = 58/64$).  It is clear that any two binary characters on a set of size three must be compatible, since they are (both) trivial, in agreement with $p_3=1$.   For a set of size four, there are $2^4 \times 2^4$ ordered pairs of binary characters $(c_1, c_2)$ and precisely $6 \times 2^2 = 24$ pairs are incompatible, which gives the proportion of pairs that are compatible as $1-24/16^2 = 58/64=p_4$. 

{\em Proof of Proposition~\ref{helps}}

Parts (a) and (c) rely on the following observation: Recall that a binary character $\chi$ is compatible with a phylogenetic $X$-tree if either $T$ or some refinement of $T$ displays $\chi$.
Now, any refinement of $T$ that displays $\chi$ is achieved by resolving a uniquely determined  single interior vertex of $T$ (\cite{sem03}, Lemma 3.1.7), as illustrated in Fig. 1 (this fact  is the basis of the `tree-popping' algorithm of \cite{mea83}).   Now, if an interior vertex $v$ has degree $d$ then there are $2^{d}$ ways to select a subset $S(v)$ of the incident branches.  For each such selection we obtain a binary character that is compatible with $T$, defined by the condition that the state of each leaf is 1 if and only if the path from that leaf to $v$ contains an edge in $S(v)$. Moreover each binary characters that is compatible with $T$ can be generated in this way (for some interior vertex $v$ and some set $S(v)$).
  Thus, the total number of binary characters on $X$ that are 
compatible with $T$ is at most $\sum_{v \in I(T)} 2^{{\rm deg}(v)}$, and since there are $2^n$ binary characters on $X$ in total, then the probability that a random character is compatible with $T$ is as bounded in Part (c).  

To establish Part (a) we view a binary character $\chi$ that divides $X$ into blocks of size $r$ and $n-r$ as  a phylogenetic $X$-tree $T$ with two interior vertices $v_1$ and $v_2$ of degrees $r+1$ and $n-r+1$ (Fig. 1 provides an example with $n=8, r=3$).  Let $E_i$ be the leaf edges of $T$ adjacent to $v_i$ (for $i=1,2$) and let $e$ be the edge $\{v_1, v_2\}$.  
Recalling
the definition of $S(v)$ from the previous paragraph, the total number of binary characters on $X$ that are compatible with $T$ is the $2^{r+1}$ choices for  $S(v_1)$ plus the $2^{n-r+1}$ choices for $S(v_2)$ minus the four choices that are counted twice as they produce the same character, namely:
 \begin{itemize}
\item[(i)]  $S(v_1) = S(v_2)=\emptyset$; 
\item[(ii)] $S(v_1) = E_1, S(v_2)= \{e\}$;
\item[(iii)] $S(v_1) = \{e\}, S(v_2)= E_2$; and
\item[(iv)] $S(v_1) = E_1 \cup \{e\}, S(v_2)= E_2 \cup \{e\}$.
\end{itemize}
 Thus the number of characters compatible with $\chi$ is $2^{r+1} + 2^{n-r+1}-4$, and  dividing by $2^n$ (the total number of binary characters on $X$) gives the expression in Part (a).

{\em Part (b):} Let $p_n(r)$ be the probability described by Part (a) for all values of $r$ between 0 and $n$, and
let $\pi_n(r) = \binom{n}{r}\left(\frac{1}{2}\right)^n$ be the probability that a random binary character divides
$X$ into blocks of size $r$ and $n-r$.  Then, by the law of total probability,  the value of $p_n$ defined in Part (b) can be written as 
\begin{equation}
\label{pneq}
p_n = \sum_{r=0}^n \pi_n(r) p_n(r).
\end{equation}
Now, since $\pi_n(0)p_n(0)=  \pi_n(n)p_n(n) = \left(\frac{1}{2}\right)^n$,  and we can write
the term on the right of Eqn.~(\ref{pneq})  as:
$\left(\frac{1}{2}\right)^{n-1} + \sum_{r=1}^{n-1} \pi_n(r) p_n(r)$.
Moreover,  from Part (a), for $1\leq r\leq n$ we can replace $p_n(r)$  by  $\alpha_n(r) = \left(\frac{1}{2}\right)^{r-1} + \left(\frac{1}{2}\right)^{n-r-1} - \left(\frac{1}{2}\right)^{n-2}$ to get:
$p_n= \left(\frac{1}{2}\right)^{n-1} + \sum_{r=1}^{n-1} \pi_n(r) \alpha_n(r).$
By a slight rearrangement, this is equivalent  to:
$$p_n= \left(\frac{1}{2}\right)^{n-1} + \left(\frac{1}{2}\right)^{n}\left[ - \alpha_n(0)- \alpha_n(n)+ \sum_{r=0}^n \binom{n}{r}\alpha_n(r) \right].$$
From here, use of the identity: $\sum_{r=0}^n \binom{n}{r} x^r = (1+x)^r$ for $x=\frac{1}{2}$ and straightforward, if tedious algebra, leads to the given expression for $p_n$.

{\em Part (d):}  If $T$ has $k$ interior edges, then $T$ has $|I(T)|= k+1$ interior vertices (since the total number of edges is $n+k$, and the total number of vertices is $n+|I(T)|$ and
since $T$ is a tree, the number of vertices exceeds the number of edges by exactly 1). 
Moreover, by the degree-sum formula,   $\sum_{v \in I(T)} {\rm deg}(v) +n\cdot 1$  is twice the number of edges of $T$ (i.e. $2(n+k)$) and so:
\begin{equation}
\label{sumeq}
\sum_{v \in I(T)} {\rm deg}(v)= n+2k, \mbox{ with } 3 \leq {\rm deg}(v)\leq n-k \mbox{ for all } v\in I(T).
\end{equation}
If we place the interior vertices in an arbitrary order $v_1, \ldots, v_{k+1}$ and let $x_i:= n-{\rm deg}(v_i)$, then from Part (c) we have:
\begin{equation}
\label{pT}
p_T \leq \sum_{i=1}^{k+1} \left(\frac{1}{2}\right)^{x_i}.
\end{equation}
Now, the constraints in (\ref{sumeq}) are equivalent to the following constraints on the $x_i$: 
 $$k \leq x_i \leq n-3 \mbox{ and } \sum_{i=1}^k x_i = k(n-2),$$ 
and the right-hand-side of (\ref{pT}) is maximised under these constraints when 
$x_i=n-3$ for $k$ values of $i$, and $x_i = 3$ for the one remaining value of $i$. This leads to the expression in (d). 
\hfill$\Box$

\subsection{Proof Theorem~\ref{mainthm}}

We first note that it suffices to prove the results under the assumption that no unvaried characters are removed.
For Parts (i) and (ii) this follows from the observation that removing unvaried random characters from $S_2$ cannot increase
the probability of either (i) a pair of characters in $S_2$ being compatible, or (ii) a character from $S_2$ being compatible with every non-trivial character from $S_1$.
For Part (iii) the probability that a random binary character on $n$ taxa is  unvaried is $\frac{1}{2^{n-1}}$; this ensures that the same stated results apply (asymptotically) for large $n$.

\mbox{ }

{\em Part (i): }  By Boole's inequality, the probability that at least one pair of the $M$ random binary characters in $S_2$  are compatible is bounded above by $\binom{M}{2}p_n$, and this is less or equal to $\binom{M}{2} \cdot \left(\frac{3}{4}\right)^n \leq 2M^2 \left(\frac{3}{4}\right)^n$ by Part (a) of Proposition~\ref{helps}.   Thus, if this last quantity is at most $\epsilon$ then 
with probability at least $1-\epsilon$ no pair of the random binary characters in $S_2$ will be compatible.  The remainder of the first two claims in part (i) now follows by simple combinatorics, noting that the  characters in $S_1$ are all compatible with each other. The proof of the third claim in part (i) was described immediately following the theorem.

{\em Part (ii): }  A character $\chi$ in $S_2$ is compatible with all of the characters in $S_1$, precisely if $\chi$ is compatible with the minimal
phylogenetic $X$-tree $T$ that displays the $K$ distinct partitions induced by the non-trivial characters in $S_1$. Since $T$ has $K$ interior edges, we can apply Part (d) of Proposition~\ref{helps}  
to conclude that the probability a random character is compatible with all of the characters in $S_1$ is at most $\left(K \left(\frac{1}{2}\right)^{n-3} + \left(\frac{1}{2}\right)^{K}\right)$.
Thus, if this quantity is at most $\delta/M$, then, again by Boole's inequality, the probability that at least one of the $M$ characters in $S_2$ is compatible with all the characters in $S_1$ 
is at most $\delta$.  This completes the proof of part (ii).

{\em Part (iii): }
For a binary character $\chi$ on $X$, and any phylogenetic $X$-tree $T$ let $ps(\chi,T)$ denote the parsimony score of $\chi$ on $T$.
For a random binary character $f$ on $X$, we let $S_T$ denote the (random variable) $ps(f, T)$.  
Moreover, if $T'$ is a binary phylogenetic $X$-tree selected uniformly at random
we let $\Delta_T$ be the (compound) random variable defined by $\Delta_T: = S_T - S_{T'}$.  Notice that $\Delta_T$ has two sources of randomness -- firstly the choice of $T'$, and once this is given, $S_T-S_{T'}$  then has variation due to the random character $f$.  We first establish the following result (where, in Lemma~\ref{smart}, the `outer' expectation in (i) and probability in (ii) is with respect to the random tree $T'$, while the `inner' (conditional) variance in (i) and (ii) is with respect to the random character (conditional on the choice of random tree $T'$): 

\begin{lemma}
\label{smart}
\mbox{ }

{\rm
 \begin{itemize}
 \item[(i)] $\EE[{\rm Var}[\Delta_T|T']]= {\rm Var}[\Delta_T] \geq \left(\frac{4}{27}-o(1) \right) n,$
 \item[(ii)]  $\PP\left({\rm Var}[\Delta_T |T'] \geq \frac{2}{27}n\right) \geq \frac{1}{3} -o(1),$
 \end{itemize}
 where $o(1)$ refers here and throughout the rest of the paper to any term that decays to zero as $n$ grows. 
 }
\end{lemma}
Our proof of this lemma combines a number of ideas. One is the well-known variance formula that applies when a random variable $Y$ depends on another random variable $W$:
\begin{equation}
\label{tele}
{\rm Var}[Y] = {\rm Var}[\EE[Y|W]] + \EE[{\rm Var}[Y|W]],
\end{equation}
Also helpful is the fact that the distribution of the parsimony score of a random binary character is the same across all binary trees \citep{ste93}, which implies that
 $\EE[ps(f, T_a)-ps(f, T_b)] = 0,$ for all pairs of binary phylogenetic $X$-trees $T_a, T_b$. So,  by taking $T_a=T$ and $T_b$ the random binary tree $T'$, we have
$\EE[\Delta_T|T'] = 0$. This in turn implies that  $\EE[\Delta_T] = \EE[\EE[\Delta_T|T'] ] = 0$, and that
${\rm Var}[\EE[\Delta_T|T']] =0$, which further implies, by Eqn.~(\ref{tele}) (using $Y=\Delta_T, W=T'$) that 
${\rm Var}[\Delta_T]  = 0+ \EE[{\rm Var}[\Delta_T|T']],$
establishing the first part (the equality) in Lemma~\ref{smart}(i). 

To establish the inequality in Lemma~\ref{smart}(i)  we make a further crucial observation.  Let $R$ be the random variable that records how many leaves in the generated random binary character $f$ are in state 1.   By Eqn. (\ref{tele}) (applied with $Y=\Delta_T$, $W=R$) we  have
\begin{equation}
\label{tele2}
{\rm Var}[\Delta_T] \geq \EE[{\rm Var}[\Delta_T | R]],
\end{equation}
 (since the second first term on the right-hand-side of Eqn. (\ref{tele}) is always non-negative).

Furthermore,  although the random variables $S_T$ and $S_{T'}$ are dependent, once we condition on $R$ they become
(conditionally) independent.   Therefore,
\begin{equation}
\label{vaRR}
{\rm Var}[\Delta_T|R] ={\rm Var}[S_T|R] + {\rm Var}[S_{T'}|R]
\end{equation}
Moreover, both the expressions on the right of this last equation are continuous functions of $p = R/n$ of the form 
$\varphi_T(p)n$ and $\varphi(p)n$, respectively; moreover, for any binary tree $T$, Corollary 7.2 of \cite{ste93} gives:
$$\lim_{ p \rightarrow \frac{1}{2}} \varphi_T(p) = \frac{2}{27} - o(1),$$
while Theorem 1 of  \cite{moo93} gives:
$$\mbox{ and } \lim_{ p \rightarrow \frac{1}{2}} \varphi(p)= \frac{2}{27} - o(1)$$ 
(recall $o(1)$ denote any term that converge to 0 as $n$ grows). 
By  the law of large numbers (or the Central Limit Theorem), $p$ converges in probability to $\frac{1}{2}$ as $n$ grows, and so by Eqn.(\ref{tele2}):
\begin{equation}
\label{helps3}
{\rm Var}[\Delta_T] \geq \EE[{\rm Var}[\Delta_T | R]] = \left( \frac{4}{27} -o(1)\right)n.
\end{equation} 
which establishes the inequality in Lemma~\ref{smart}(i).

We now use Part (i) of Lemma~\ref{smart} to derive Part (ii), by means of the following general observations. 
Suppose that $a,b>0$ and that $Y$ is any random variable that always lies between
$0$ and $a(2+\delta)$ and that $\EE[Y] \geq a(1-\epsilon)$.  This ensures the inequality: $$a(1-\epsilon)\leq \EE[Y] \leq a(2+\delta) \PP(Y \geq a/2) + (a/2)(1-\PP(Y \geq a/2)),$$ which simplifies to:
\begin{equation}
\label{ppineq}
\PP(Y \geq a/2) \geq \frac{1}{3}\cdot \frac{1-  2\epsilon}{1+2\delta/3}.
\end{equation}
We apply Inequality~(\ref{ppineq}) with $Y = {\rm Var}(\Delta_T |T')$, so that $\EE[Y] ={\rm Var}(\Delta_T)$, and taking $a= \frac{4}{27}n$,
Lemma~\ref{smart}(i)  ensures that  $\EE[Y] \geq a(1-\epsilon),$
 where and $\epsilon$ is a term of order $o(1)$.
It remains to show that $Y$ is bounded above by $a(2+\delta)$ where $\delta$ is also a  term of order $o(1)$.

By definition, ${\rm Var}(\Delta_T|T') ={\rm Var}(S_T|T')+ {\rm Var}(S_{T'}|T') -2{\rm Cov}(S_T, S_{T'}|T')$
and ${\rm Var}(S_T|T') = {\rm Var}(S_T)$ and ${\rm Var}(S_{T'}|T')$ are both equal to $\left(\frac{2}{27} + o(1)\right)n$ \citep{ste93}.
Moreover, $|Cov(S_T, S|T')|$ is at most the square root of the product of these two variances.  
In summary, ${\rm Var}(\Delta_T|T')  \leq a(2+o(1))$.
Inequality (\ref{ppineq})  now gives Lemma~\ref{smart}(ii).

Returning to the proof of Part (iii) of the Theorem~\ref{mainthm},   Theorem 4 of \citep{bry09},  shows that the proportion of binary trees that share a given number of non-trivial splits with any given binary tree $T$ on the same leaf set is asymptotically Poisson with a mean $\lambda_T$ equal to the number of `cherries' of $T$ divided by $2n$. Since the number of cherries in any binary tree with $n$ leaves  is at most $\frac{n}{2}$, we have $\lambda_T \leq \frac{1}{4}$. Thus the probability that $T'$ shares no non-trivial splits with $T_1$ is at least $\exp(-1/4) \approx 0.78$.  Since all the characters in $S_1$ are displayed by $T_1$, the corresponding splits of the non-trivial characters constitute a subset of the splits of $T_1$, and so the probability that $T'$ displays one or more  non-trivial character in $S_1$ is also at most $\exp(-1/4)$. 

We combine this with Lemma~\ref{smart}(ii) which  shows that for $n$ large, with probability at least $\sim \frac{1}{3}$, a binary phylogenetic $X$-tree $T'$ selected uniformly at random will
have ${\rm Var}(\Delta_T|T')$ at least $\frac{2}{27}n$.   Thus, the probability that a  binary phylogenetic $X$-tree $T'$ selected uniformly at random satisfies both (a) ${\rm Var}(\Delta_T|T')$ at least $\frac{2}{27}n$, and (b) $T'$ displays none of the
non-trivial characters in $S_1$ is (asymptotically) at least $1- (1-\exp(-1/4)+ (1-1/3) ) > 0.11$, by the Bonferroni inequality.

For the rest of the proof $T'$ will refer to any one of these (at least) $11\%$ of all unrooted binary phylogenetic $X$-trees that satisfy properties (a) and (b) in the last sentence.
Let $ps(D, T_1)$ and $ps(D, T')$ be the parsimony score of the data $D$ (consisting of the characters
in $S_1$ and $S_2$ intertwined arbitrarily) on $T_1$ and on $T'$, respectively.
Let $\delta P = ps(D,T_1)-ps(D, T')$, the difference in the parsimony score of $T_1$ and $T'$ for the data $D$.

Notice that we can write $\delta P = \delta P_1 +\delta P_2$ where $\delta P_i$ is the parsimony score difference of the two trees (for $T_1$ minus that for $T'$) on the characters in $S_i$.
Now,  $P_2$ is a sum of $M$ independent and identically distributed random variables, each with expected value 0, and
finite variance that is at least $c=\frac{2}{27}n$.  Thus, by the Central Limit Theorem, $\delta P_2/\sqrt{cnM}$ (asymptotically with $M$) follows a standard normal distribution
with mean 0 and standard deviation at least 1. Moreover, if $t$ is the number of characters in $S_1$ that are autapomorphic  (i.e. trivial but not unvaried) then any binary tree (in particular $T'$) satisfies  $ps(S_1, T')  \leq Lm+t$, and (by the assumption that $T$ is compatible with all the characters in 
$T_1$),  $ps(S_1, T_1) = m+t$. Thus  $\delta P_1 = ps(S_1, T_1)-ps(S_1, T')  \geq -m(L-1).$

Now $T'$ has a lower parsimony score than $T_1$ precisely if $\delta P>0$ and the probability of this event is given by:
$$\PP(\delta P_2 > -\delta P_1) \geq \PP(\delta P_2 > m(L-1)) \geq  \PP\left(Z > \frac{m(L-1)}{\sqrt{cnM}}\right),$$
where $Z$ is a standard normal random variable.  So, if $\frac{m(L-1)}{\sqrt{cnM}}\leq 1$, then  $\PP(\delta P>0) \geq \PP(Z> 1) \geq 0.15$.

Consequently, the probability $P$ that a binary phylogenetic $X$-tree $T'$ selected uniformly at random is  both more parsimonious for $D$ than $T_1$ and  displays none of the non-trivial characters in $S_1$ is at least $P'Q$, where $P'$ is the probability that $T'$ satisfies both (a) ${\rm Var}(\Delta_T|T')$ at least $\frac{2}{27}n$, and (b) $T'$ displays none of the
non-trivial characters in $S_1$, while  $Q$ is the conditional probability that such a tree $T'$ (that satisfies (a) and (b)) is more parsimonious for $D$ than $T$.
We have shown that $P > 0.11$ and $Q > 0.15$ so  $P \geq P'Q > 0.0165$.
Now, the expected number of binary phylogenetic $X$-trees that are more parsimonious than $T_1$ on $D$ and  displays none of the non-trivial characters in $S_1$ is 
$P\cdot b(n)$ where $b(n)$ is the total number of unrooted binary phylogenetic trees with $n$ leaves.   Since $b(n) \sim \frac{1}{\sqrt{2}}\left(\frac{2}{e}\right)^{n-1} n^{n-2}$ \citep{sem03},
and $P\geq 0.0165$,  we have $P \cdot b(n) \geq 10^n$ for $n\geq 30$, as claimed.

\hfill$\Box$

\end{document}